\documentclass[12pt]{article}
\usepackage{feynmp}
\usepackage{fullpage}
\newcommand {\be} {\begin{equation}}
\newcommand{\ee} {\end{equation}}
\newcommand{\bea}{\begin{eqnarray}}
\newcommand{\eea}{\end{eqnarray}}
\newcommand{\bean}{\begin{eqnarray*}}
\newcommand{\eean}{\end{eqnarray*}}
\newcommand{\noi}{\noindent}
\begin{document}

\begin{titlepage}

\rightline{Alberta Thy 03-00 }
\rightline{hep-th/0003137}
\rightline{March 2000}

\vskip .3in 

\begin{center}
{\large{\bf Noncommutative Field Theory and Spontaneous 
Symmetry Breaking}}
\end{center}

\vskip .2in

\begin{center}
Bruce A. Campbell and Kirk Kaminsky\\ 

\vskip .1in

{\it Department of Physics, University of 
Alberta}  \\ {\it  Edmonton, Alberta, Canada T6G 2J1} \\
\end{center}

\vskip .2in

\begin{abstract}
\noi
We investigate the noncommutative analogue of the spontaneously broken linear 
sigma model at the one-loop quantum level.  In the commutative case, 
renormalization of a theory with a spontaneously broken continuous global
symmetry depends on cancellations that enable the limited set of 
counterterms consistent with that symmetry to remove the divergences
even after its spontaneous breaking, while preserving the masslessness 
of the associated Goldstone modes.  In the noncommutative case, we 
find that these cancellations are violated, and the renormalized 
one-loop correction to the inverse pion propagator explicitly yields a mass
shift which depends on the ultraviolet cutoff.  Thus, we cannot naively take 
the ultraviolet cutoff to infinity first, and then take the 
external momentum to zero to verify Nambu-Goldstone symmetry realization.
However, from the Wilsonian perspective where the cutoff is fixed and 
physical, the zero external momentum limit of the inverse pion propagator 
still vanishes, and implies the masslessness of the pion fields at one-loop.
This is another demonstration of the failure of ultraviolet and infrared 
limits to commute in noncommutative field theories, and signals the 
incompatibility of Nambu-Goldstone symmetry realization with the 
continuum renormalization of these theories.
\end{abstract}

\vskip .1in

\end{titlepage}

\newpage
\section{Introduction}


In this paper we undertake a perturbative analysis of quantum field
theories on noncommutative ${\cal R}^{n}$ which exhibit the spontaneous 
breaking of a continuous global symmetry.  The perturbative expansion of
noncommutative field theories has been the subject of much recent investigation
(see for example references [1-23]).  In the original work of Filk \cite{f},
it was suggested that despite being (infinitely) nonlocal, 
these theories exhibit the same divergence structure as their 
commutative counterparts (due to nonplanar oscillatory damping).  More 
recently, Minwalla, Van Raamsdonk, and Seiberg \cite{mrs} have shown that the 
effective cutoff of nonplanar graphs at one-loop in scalar theories replaces 
a UV divergence (which would ordinarily be cancelled with a counterterm) with 
an IR divergence in the external momenta:
\be
\Lambda_{eff}^{2} = {1\over {1\over 
\Lambda^{2}} + p \circ p } \rightarrow {1\over p \circ p } 
\hspace{10pt} , \hspace{10pt} \Lambda^{2} \rightarrow \infty    
\ee
%
They find that UV and IR limits do not commute, and suggest that UV 
divergences that persist at higher orders can be interpreted as IR 
divergences.

We continue these investigations, and examine global spontaneous 
symmetry breaking in the simplest case of the (noncommutative) linear 
sigma model.  In particular, we wish to investigate the status of 
Goldstone's theorem at one-loop for the noncommutative case.  As is 
well known, the renormalizability of spontaneously broken theories is 
more subtle because the number of counterterm vertices exceeds the 
number of renormalization parameters (eight and three respectively in the 
linear sigma model).  In the standard commutative case, the renormalizability 
of the theory, and the persistence of Goldstone's theorem ensuring the 
masslessness of the classical pions at the quantum level, involve a 
delicate cancellation between the relevant graphs and the pion 
propagator counterterm at zero external momentum [24-27].  
The pion propagator counterterm is however fixed by the sigma tadpole 
counterterm, which in turn is fixed by the usual renormalization condition 
imposed on tadpoles: namely that they vanish so that the vev of the sigma 
is not renormalized.  Thus, the persistence of masslessness of the pions 
(Goldstone's theorem) at the quantum level is a genuine prediction 
of the quantum field theory.  We now generalize this standard textbook 
calculation to the noncommutative case.

We will find that the cancellation in the calculation of the pion mass 
renormalization is violated already at one loop order, and the 1PI 
{\it renormalized} (ie. after adding the counterterm 
fixed by the tadpole renormalization condition) effective action depends 
explicitly on the UV cutoff $\Lambda$, so the naive continuum limit does
not exist; there is no more counterterm freedom to subtract off these 
divergences.  Put another way, turning on the noncommutativity parameter 
induces UV cutoff dependence in the one-loop corrections to the renormalized 
pion propagator, and renders renormalization inconsistent with Goldstone
symmetry realization.  Viewed however as a Wilsonian effective theory where 
$\Lambda$ is fixed, the limit of the mass correction as the external 
momentum is taken to zero still vanishes, so that Goldstone's theorem is 
satisfied for the Wilsonian action.  The difference between the Wilsonian
effective theory, and the putative continuum renormalized theory is a direct
consequence of the noncommutativity of the UV and IR limits in noncommutative
field theories.

The basic idea behind this rather elementary calculation is that the 
noncommutativity does {\it not} affect the tadpole 
calculation as no external momentum flows into a trilinear tadpole vertex.  
Insofar as the sigma tadpole measures quantum corrections to the sigma vev, 
this suggests that the order parameter for spontaneous symmetry breaking is
insensitive to the underlying noncommutativity. Furthermore, since the 
tadpole counterterm, and the pion propagator counterterm are 
(essentially) the same (modulo the momentum dependent wavefunction 
renormalization parameter), the latter is therefore unmodified, and 
{\it fixed} with respect to renormalization.  Now however, the re-weighting 
of planar graphs (due to noncommutativity) with respect to commutative graphs, 
and the distinct behaviour of new nonplanar graphs occurring in the 
one-loop contributions to the pion propagator, lead to an 
inexact UV cutoff cancellation with the associated counterterm.  We emphasize 
that this cannot be evaded by simply imposing that the propagator 
counterterm cancel the cutoff dependence of the one-loop planar and nonplanar 
diagrams (in effect changing the renormalization scheme), because now the 
sigma tadpole corrections will unavoidably diverge.  This is a direct 
consequence of the relationship between the counterterm vertices in a 
spontaneously broken quantum field theory.  In the following we calculate
the one-loop renormalization of the inverse pion propagator in the linear 
sigma model of noncommutative field theory, and demonstrate the 
incompatibility of continuum renormalization with spontaneous symmetry 
breaking.

\section{Background and Formalism}

In this section we introduce the noncommutative geometry and linear 
sigma model formalism we will need for the calculations of the next 
section in a reasonably self-contained way.  The underlying noncommutative
${\cal R}^{n}$ is labelled by $n$ coordinates satisfying
\be
 \left[ {\hat x}^{\mu},{\hat x}^{\nu} \right] = i \theta^{\mu\nu} 
 \hspace{10pt} , \hspace{10pt} \theta^{\mu\nu} \in {\cal C}
\label{fund-comm}
\ee
%
so that the commutators are C-numbers.  Let ${\cal A}_{x}$ be the 
associative algebraic structure naturally defined by (\ref{fund-comm}).  
We are naturally interested in functions defined over the  
noncommutative spacetime.  Following Weyl \cite{w} define
\be
O(f) = {1\over (2 \pi)^{n/2}} \int d^{n} k \hspace{5pt} e^{i k_{\mu} 
{\hat x}^{\mu}} \hspace{5pt} \tilde{f}(k)
\ee
%
where $\tilde{f}$ is the Fourier transform of $f$:
\be
\tilde{f}(k) = {1\over (2 \pi)^{n/2}} \int d^{n} x \hspace{5pt} e^{-i k_{\mu} x^{\mu}} 
\hspace{5pt} f(x)
\ee
%
This uniquely associates an operator in $O(f) \in {\cal A}_{x}$ with a function
of classical variables: replacing the commuting variables $x$ with 
operators ${\hat x}$ in a symmetric fashion.  The product of two 
such operators is then defined in the obvious way
\bea
O(f) \cdot O(g) &=& {1\over (2 \pi)^{n}} \int d^{n}k d^{n}p \hspace{5pt} 
e^{i k_{\mu} {\hat x}^{\mu}} e^{i p_{\nu} {\hat x}^{\nu}} \hspace{5pt} 
\tilde{f}(k) \tilde{g}(p) \nonumber \\
&=& {1\over (2 \pi)^{n}} \int d^{n}k d^{n}p \hspace{5pt} e^{i (k_{\mu} + 
p_{\mu}) {\hat x}^{\mu} - {i\over2} k_{\mu} \theta^{\mu\nu} p_{\nu}} \hspace{5pt} 
\tilde{f}(k) \tilde{g}(p)
\eea
%
where on the second line we have used the Baker-Campbell-Hausdorff lemma 
$e^{A} e^{B} = e^{A+B-1/2[A,B]+\ldots}$, and the fact that for the canonical 
structure (\ref{fund-comm}), the higher commutators vanishes.  This 
allows us to establish a homomorphism, $O(f)\cdot O(g) = O(f\ast g)$, between 
this operator product ($\cdot$) and the Moyal \cite{m} product ($\ast$) of 
ordinary functions:
\bea
(f \ast g)(x) &=& {1\over (2 \pi)^{n}} \int d^{n}k d^{n}p \hspace{5pt} e^{i 
(k_{\mu} + p_{\mu}) x^{\mu}} e^{- {i\over2} k_{\mu} \theta^{\mu\nu} p_{\nu}} 
\hspace{5pt} \tilde{f}(k) \tilde{g}(p) \nonumber \\
&=& e^{+{i\over 2} \theta^{\mu\nu} {\partial\over \partial y^{\mu}} 
{\partial\over \partial z^{\nu}} } f(y) g(z) |_{y,z\rightarrow x}     
\label{Moyal}
\eea
%
This induced homomorphism allows us to view the algebra of functions 
on noncommutative ${\cal R}^{n}$ as the algebra of ordinary functions 
on commutative ${\cal R}^{n}$ with the Moyal $\ast$-product instead 
of the usual pointwise product.  In particular, we can study field 
theories defined by classical actions of the usual form $S = \int 
d^{n}x {\cal L}[\phi]$ but with the $\ast$-product of fields.

Now consider the spontaneously broken linear sigma model.  At this point 
we specialize to four dimensions for the remainder of this paper.  
Furthermore, our sigma-model conventions will be essentially those of 
Peskin and Schroeder \cite{ps}.  The commutative linear sigma model involves 
a set of N interacting real scalar fields $\phi^{i}(x)$ with a continuous 
$O(N)$ internal symmetry.  Renormalizability (of the commutative case) in four 
dimensions implies the Lagrangian:
\be
{\cal L} = {1\over 2} \partial_{\mu} \phi^{i} \partial^{\mu} \phi^{i}
+ {1\over 2} \mu^{2} (\phi^{i})^{2} - {\lambda\over 4} \left[ 
(\phi^{i})^{2} \right]^{2}       
\label{sym-Lag}
\ee
%
with implicit sums over the internal i index.  Note the rescaling of 
the coupling $\lambda$ to avoid awkward factors of $1\over 6$.  The 
{\it global} $O(N)$ symmetry acts as
\be
\phi^{i} \rightarrow R^{ij} \phi^{j}  
\ee
%
where $R$ is a spacetime-{\it constant} $N\times N$ orthogonal matrix.
Because the symmetry is global (so $R$ is constant), the 
noncommutative generalization of this symmetry will manifestly pose 
no problem with respect to the Moyal product, which explicitly 
degenerates into the pointwise product if one of the factors is a 
spacetime constant.  In particular we will not need to worry about the 
intricacies of the noncommutative generalization of gauge invariance 
(see for example \cite{mssw}).

For $\mu^{2}>0$, the classical potential is minimized by a {\it 
constant} field $\phi^{i}_{0}$ configuration such that
\be
(\phi^{i}_{0})^{2} = {\mu^{2}\over \lambda}
\ee
%
and the $O(N)$ symmetry is spontaneously broken.  Since this 
determines only the length of the vector $\phi^{i}$, the rotational 
invariance allows us to choose coordinates so that
\be
\phi^{i}_{0} = (0,0,\ldots,0,v), \qquad v\equiv 
{\mu\over\sqrt{\lambda}}
\ee
%
Then define $\sigma = \phi_{0}^{n} - v$ such that
\be
\phi^{i} = (\pi^{k},\sigma), \qquad k=1,..,N-1
\ee
%
with $<\sigma>=0$.  The linear sigma-model Lagrangian (\ref{sym-Lag}) 
rewritten in terms of these fields, and Wick rotated to Euclidean 
space (via $x^{0}=-i x^{0}_{E}$) yields the following (commutative) action:
\bea
S_{E} &=& - \int d^{4}x \hspace{5pt} \left[ {1\over 2} \partial_{\mu} 
\pi^{k} \partial^{\mu} \pi^{k} + {1\over 2} \partial_{\mu} \sigma 
\partial^{\mu} \sigma + {1\over 2} (2\mu^{2}) \sigma^{2} + 
{\lambda \over 4} [(\pi^{k})^{2}]^{2}  \right. \nonumber \\ 
& & + \left. {\lambda\over 4} \sigma^{4} + v\lambda\sigma(\pi^{k})^{2} + 
{\lambda\over 2} (\pi^{k})^{2} \sigma^{2} + \lambda v \sigma^{3} 
\right]
\eea
%
which reveals explicitly the masslessness of the $N-1$ pions of 
Goldstone's theorem at tree level.  The renormalization counterterm 
structure determined from the symmetric theory (if the $O(N)$ 
symmetry is to hold quantum mechanically) is given by:
\bea
-{\cal L}_{E,ct} &=& {\delta_{Z}\over 2} \partial_{\mu} \phi^{i} 
\partial^{\mu} \phi^{i} + {1\over 2} \delta_{\mu} (\phi^{i})^{2} + 
{\delta_{\lambda}\over 4} [(\phi^{i})^{2}]^{2}  \nonumber \\
&=& {\delta_{Z}\over 2} (\partial_{\mu} \pi^{k})^{2} + {\delta_{Z}\over 
2} (\partial_{\mu} \sigma)^{2} + {1\over 2}(\delta_{\mu}+ 
\delta_{\lambda} v^{2}) (\pi^{k})^{2} \nonumber \\
& & + {1\over 2} (\delta_{\mu} + 3\delta_{\lambda}v^{2})\sigma^{2} +
(\delta_{\mu} v +\delta_{\lambda}v^{3}) \sigma + 
\delta_{\lambda} v \sigma (\pi^{k})^{2} + \delta_{\lambda} v \sigma^{3}
\nonumber \\ 
& & + {\delta_{\lambda}\over 4} [(\pi^{k})^2]^{2} + {\delta_{\lambda} \over 2} 
\sigma^{2} (\pi^{k})^{2} + {\delta_{\lambda}\over 4} \sigma^{4}          
\label{counterterm-Lag}
\eea
%
Now consider the noncommutative generalization of this theory.  As 
discussed above, the effect at the Lagrangian level of the 
noncommutativity is to replace the (implicit) pointwise product of 
fields with the Moyal $\ast$-product.  However, under the spacetime integral, 
an elementary but important result is that the quadratic part of the 
action is identical with commutative theory because $\theta^{\mu\nu}$ 
is antisymmetric.  For example from (\ref{Moyal}),
\bea
(\phi_{1}  * \phi_{2} )(x) &=& \phi_{1}(x) \phi_{2}(x) + 
{i\over 2} \theta^{\mu\nu} \partial_{\mu} \phi_{1}(x) 
\partial_{\nu} \phi_{2}(x) + \ldots \nonumber \\     
&=& \phi_{1}(x) \phi_{2}(x) + {i\over 2} \theta^{\mu\nu} \left[ 
\partial_{\mu} (\phi_{1}(x) \partial_{\nu} \phi_{2} (x)) - 
\phi_{1}(x) \partial_{\mu}\partial_{\nu} \phi_{2}(x) \right] + 
\ldots  \nonumber \\
&=& \phi_{1}(x) \phi_{2}(x) + \hspace{5pt} \textnormal{total derivative}
+ \ldots  
\eea
%
where $\ldots$ represent the higher terms in $\theta$ that behave 
identically with respect to this calculation.  Thus dropping total 
derivatives, by assuming appropriate asymptotic conditions on the 
fields, we have:
\bea
& & \int d^{4}x \hspace{5pt} \phi \ast \phi = \int d^{4} x \hspace{5pt} 
\phi \cdot \phi  \nonumber \\
& & \int d^{4}x \hspace{5pt} \partial \phi \ast \partial \phi = \int d^{4} 
x \hspace{5pt} \partial \phi \cdot \partial \phi   
\eea
%
Interactions in higher powers of the fields are modified, and yield 
nontrivial phase factors in the momentum space Feynman rules, as we will see 
in a moment.  There are two possible orderings for the noncommutative 
generalizations of the quartic terms $\pi^{k} \pi^{k} \pi^{l} \pi^{l}$, and 
$\pi^{k} \pi^{k} \sigma \sigma$, so we will include both orderings for each 
interaction in the model with unit total weighting.  Thus the most general 
noncommutative, spontaneously broken linear sigma model action in Euclidean 
space:
\bea
S_{E,nc} &=& - \int d^{4}x \hspace{5pt} \left[ {1\over 2} \partial_{\mu} 
\pi^{k} \partial^{\mu} \pi^{k} + {1\over 2} \partial_{\mu} \sigma 
\partial^{\mu} \sigma + {1\over 2} (2\mu^{2}) \sigma^{2}  \right. \nonumber \\ 
& & + {\lambda \over 4} f (\pi^{k} \ast \pi^{k}) \ast ( \pi^{l} \ast \pi^{l}) 
+ {\lambda \over 4} (1-f) (\pi^{k} \ast \pi^{l}) \ast ( \pi^{k} \ast \pi^{l})
\nonumber \\ & &  + {\lambda\over 4} \sigma \ast \sigma \ast \sigma \ast 
\sigma  + v\lambda\sigma \ast (\pi^{k} \ast \pi^{k} ) + 
{\lambda\over 2} f ( \pi^{k} \ast \pi^{k} ) \ast \sigma \ast \sigma \nonumber
\\ & & \left. + {\lambda\over 2} \left( 1-f \right) (\pi^{k} \ast \sigma) \ast 
(\pi^{k} \ast \sigma) + \lambda v \sigma \ast \sigma \ast \sigma  
\right]
\label{ncsblsm}
\eea
%
and similarly for the counterterm structure.  Note that the global $O(N)$ 
symmetry of the model implies that only a single $f$
can occur.  Since we wish to examine the theory at the quantum level, we will 
need the Feynman rules.  In momentum space the Moyal product (\ref{Moyal}) of 
n fields yields
\be
\int d^{4}x \phi_{1}(x) \ast \ldots \ast \phi_{n}(x) = 
\int \left( \prod_{i=1}^{n} d^{4}p_{i} \right)  \delta\left(\sum_{i=1}^{n} p_{i}\right) 
e^{-{i\over 2} \sum_{i<j} p_{i} \times p_{j} } \tilde{\phi}_{1}(p_{1}) 
\cdots \tilde{\phi}_{n}(p_{n})                     
\ee
%
where $p_{i}\times p_{j} \equiv p_{i\mu}\theta^{\mu\nu} p_{j\nu}$.  
We will prove this in the appendix inductively.  Thus the only modification 
to the Feynman rules is due to the phase factor
\be
V(p_{1}\ldots p_{n}) = e^{-{i\over 2} \sum_{i<j} p_{i}\times p_{j} }  
\label{gen-phase}
\ee
%
at each vertex of $n$ fields with $p_{i}$ the momentum {\it into} the 
vertex from the ith field.  Otherwise the Feynman rules will be identical 
to the linear sigma model.  Note that the sum contains $n (n-1)/2$ 
terms, and by momentum conservation is invariant under cyclic permutations, 
but not under arbitrary permutations of the momenta.  We have both three and 
four point vertices in the spontaneously broken linear sigma model; a fact 
that will be crucial for our analysis.

To summarize we write the {\it symmetrized} Feynman rules for the 
spontaneously broken noncommutative linear sigma model using 
(\ref{ncsblsm}) and (\ref{gen-phase}).  Throughout solid lines refer to the 
sigma, and dotted lines refer to the pions.  As discussed above, the 
quadratic terms are not modified, so the propagators are not modified from the commutative case:
\begin{fmffile}{propagators}
  \bea
    \parbox{100pt}{\begin{fmfgraph*}(100,50)
       \fmfleft{w}
       \fmfright{e}
       \fmf{dashes_arrow,label=$p$}{w,e}
       \fmfv{label=$\pi^{i}$,label.angle=60}{w}
       \fmfv{label=$\pi^{j}$,label.angle=120}{e}
    \end{fmfgraph*}}  &=&  {\delta^{ij}\over p^{2}} \hspace{15pt} 
    \parbox{100pt}{\begin{fmfgraph*}(100,50)
       \fmfleft{w}
       \fmfright{e}
       \fmf{plain_arrow,label=$p$}{w,e}
       \fmfv{label=$\sigma$,label.angle=60}{w}      
    \end{fmfgraph*}}  =  {1\over p^{2} + 2\mu^{2}}  
  \eea
\end{fmffile}
%
The {\it symmetrized} vertices (all momenta flow into the vertex) of the 
theory are now given by:
\begin{fmffile}{vertices}
 \bea
  \parbox{70pt}{\begin{fmfgraph*}(50,50)
     \fmfsurround{v1,v2,v3}
     \fmfdot{c}
     \fmf{plain}{v1,c}
     \fmf{dashes}{v2,c}
     \fmf{dashes}{v3,c}
     \fmfv{label=$p_{3}$}{v1}
     \fmfv{label=$i \hspace{5pt} p_{1}$,label.angle=180}{v2}
     \fmfv{label=$j \hspace{5pt} p_{2}$}{v3}
  \end{fmfgraph*}}  &=&  -2 v \lambda \delta^{ij} \cos( {
    p_{1}\times p_{2}\over 2})
  \label{pi-pi-sigma}
 \eea
 \bea 
  \parbox{70pt}{\begin{fmfgraph*}(50,50)
     \fmfsurround{v1,v2,v3}
     \fmfdot{c}
     \fmf{plain}{v1,c}
     \fmf{plain}{v2,c}
     \fmf{plain}{v3,c}
     \fmfv{label=$p_{3}$}{v1}
     \fmfv{label=$p_{1}$}{v2}
     \fmfv{label=$p_{2}$}{v3}
  \end{fmfgraph*}}  =  - 2 v \lambda \left[\cos({p_{1}\times p_{2}\over 2})
  + \cos({p_{1}\times p_{3}\over 2}) + \cos({p_{2}\times p_{3}\over 2}) \right]
  \label{sigma-sigma-sigma}
 \eea
 \vspace{0.5\baselineskip}
 \bea
  \parbox{70pt}{\begin{fmfgraph*}(50,50)
     \fmfsurroundn{v}{8}
     \fmfdot{c}
     \fmf{dashes}{v2,c}
     \fmf{dashes}{v4,c}
     \fmf{plain}{v6,c}
     \fmf{plain}{v8,c} 
     \fmfv{label=$j \hspace{5pt} p_{2}$}{v2}
    \fmfv{label=$i \hspace{5pt} p_{1}$}{v4}
    \fmfv{label=$p_{3}$}{v6}
    \fmfv{label=$p_{4}$}{v8}
  \end{fmfgraph*}} & = & - 2\lambda \delta^{ij} \left[ f \cos({p_{1}\times 
     p_{2} \over 2}) \cos({p_{3}\times p_{4}\over 2}) + (1-f) 
     \cos({p_{1} \times p_{3}\over 2} + 
     {p_{2} \times p_{4}\over 2}) \right] \nonumber \\ & &
  \label{pi-pi-sigma-sigma} 
 \eea
 \bea
  \parbox{50pt}{\begin{fmfgraph*}(50,50)
     \fmfsurroundn{v}{8}
     \fmfdot{c}
     \fmf{plain}{v2,c}
     \fmf{plain}{v4,c} 
     \fmf{plain}{v6,c}
     \fmf{plain}{v8,c}
     \fmfv{label=$p_{2}$}{v2}
     \fmfv{label=$p_{1}$}{v4}
     \fmfv{label=$p_{3}$}{v6}
     \fmfv{label=$p_{4}$}{v8}
  \end{fmfgraph*}} & =&  - 2 \lambda \left[ \cos({p_{1}\times p_{2}\over 2}) 
    \cos({p_{3}\times p_{4}\over 2}) + \cos({p_{1}\times p_{3}\over 2}) 
    \cos({p_{2}\times p_{4}\over 2}) \right. \nonumber \\ 
    & & \left. + \cos({p_{1}\times p_{4}\over 2}) 
    \cos({p_{2}\times p_{3}\over 2}) \right]
  \label{four-sigma} 
 \eea
 \bea
  \parbox{70pt}{\begin{fmfgraph*}(50,50)
     \fmfsurroundn{v}{8}
     \fmfdot{c}
     \fmf{dashes}{v2,c}
     \fmf{dashes}{v4,c} 
     \fmf{dashes}{v6,c}
     \fmf{dashes}{v8,c}
     \fmfv{label=$j \hspace{5pt} p_{2}$}{v2}
     \fmfv{label=$i \hspace{5pt} p_{1}$}{v4}
     \fmfv{label=$k \hspace{5pt} p_{3}$}{v6}
     \fmfv{label=$l \hspace{5pt} p_{4}$}{v8}
  \end{fmfgraph*}}  &=&  -2 \lambda \left[ \delta^{ij}\delta^{kl} \left( 
     f \cos({p_{1}\times p_{2}\over 2}) \cos({p_{3}\times p_{4}\over 2})
     + (1-f) \cos( {p_{1}\times p_{3}\over 2} + {p_{2}\times p_{4}\over 2}) 
     \right) \right. \nonumber \\ & & + 
     \delta^{ik}\delta^{jl} \left( f \cos({p_{1}\times p_{3}\over 2})
     \cos({p_{2}\times p_{4}\over 2}) + (1-f) \cos({p_{1}\times p_{2}\over 2}
     + {p_{3}\times p_{4}\over 2}) \right) \nonumber \\ & & + \left.
     \delta^{il}\delta^{jk} \left( f \cos({p_{1}\times p_{4}\over 2})
     \cos({p_{2}\times p_{3}\over 2}) + (1-f) \cos({p_{1}\times p_{2}\over 2}
     + {p_{4}\times p_{3}\over 2}) \right) \right] \nonumber \\ & &  
  \label{four-pi}
 \eea
\end{fmffile}
%
Working with the totally symmetrized vertices allows us to capture both planar
and nonplanar terms at once, though to make this explicit we will 
{\it a posteriori} extract planar and nonplanar parts below.

In the calculations of the next section we will only be interested in one and 
two point amplitudes, so the counterterms from (\ref{counterterm-Lag}) we 
will need are given by
\begin{fmffile}{counterterms}
  \bea
   \parbox{90pt}{\begin{fmfgraph*}(90,50)
      \fmfkeep{pion-counterterm}
      \fmfleft{w}
      \fmfright{e}
      \fmf{dashes}{w,c,e}
      \fmfv{decor.shape=cross}{c}
      \fmfv{label=$i$,label.angle=60}{w}
      \fmfv{label=$j$,label.angle=120}{e}
   \end{fmfgraph*}} = -\delta^{ij} (\delta_{\mu} + \delta_{\lambda} 
   v^{2} - \delta_{z} p^{2} ) \hspace{10pt}      
   \parbox{50pt}{\begin{fmfgraph}(50,50)
       \fmfkeep{sigma-tadpole-counterterm}
       \fmfleft{w}
       \fmfright{e}
       \fmf{plain}{w,e}
       \fmfv{decor.shape=cross}{e} 
   \end{fmfgraph}} \hspace{10pt} = -(\delta_{\mu} v +\delta_{\lambda} v^{3})
   \nonumber    
  \eea
  \be
   \parbox{90pt}{\begin{fmfgraph}(90,50)
      \fmfleft{w}
      \fmfright{e}
      \fmf{plain}{w,c,e}
      \fmfv{decor.shape=cross}{c}
   \end{fmfgraph}} = -(\delta_{\mu}+ 3 \delta_{\lambda}v^{2} - 
   \delta_{z} p^{2})
   \label{counterterms}
  \ee
\end{fmffile}
\section{One-Loop Quantum Level Calculations}

The linear sigma model contains three counterterm parameters, 
and so three renormalization conditions are needed.  These are 
conventionally taken to be conditions specifying the field strength 
of $\sigma$, the $4$-$\sigma$ scattering amplitude at threshold, and the 
vanishing of the one-point amplitude or vacuum expectation value 
renormalization of the $\sigma$.  Everything else, including the 
masses of both $\sigma$ and the $\pi$'s, are predictions of the 
quantum field theory.  In the renormalized perturbation theory sense, 
if the counterterms can be adjusted order by order to maintain the 
renormalization conditions, and yield finite predictions for 
everything else then the theory is perturbatively renormalizable.  In 
the following we will examine the one-loop quantum structure of the 
noncommutative theory.

Of course, our results will not depend on this particular 
renormalization scheme, but since our explicit interest is in 
Goldstone's theorem at the quantum level for the noncommutative case, 
and in this scheme the masslessness of the pions is a {\it prediction} of the 
quantum field theory, we will employ it in the following.\footnote{Actually 
we will directly use only the renormalization condition on the sigma 
one-point amplitude: namely that it vanishes.  We will consider the 
general case later.}  In the ordinary commutative calculation of pion mass 
renormalization (see \cite{ps} for example) the masslessness of the 
pions at one-loop (Goldstone's theorem) follows from a cancellation of 
the following graphs at zero external momentum ($p=0$): 
\begin{fmffile}{planar}
 \bea
  \parbox{120pt}{\begin{fmfgraph*}(120,60)
    \fmfkeep{planar-1}
    \fmfleft{w}
    \fmfright{e}
    \fmf{dashes_arrow,label=$p$}{w,c,e}
    \fmffreeze
    \fmftop{n}
    \fmf{dashes_arrow,left,label=$k$}{c,n}
    \fmf{dashes,right}{c,n}
    \fmfdot{c}
    \fmfv{label=$i$,label.angle=60}{w}
    \fmfv{label=$j$,label.angle=120}{e}
  \end{fmfgraph*}} + 
  \parbox{120pt}{\begin{fmfgraph*}(120,60)
    \fmfkeep{planar-2}
    \fmfleft{w}
    \fmfright{e}
    \fmf{dashes_arrow,label=$p$}{w,c,e}
    \fmffreeze
    \fmftop{n}
    \fmf{plain_arrow,left,label=$k$}{c,n}
    \fmf{plain,right}{c,n}
    \fmfdot{c}
    \fmfv{label=$i$,label.angle=60}{w}
    \fmfv{label=$j$,label.angle=120}{e}
  \end{fmfgraph*}}
  \nonumber \\ + \quad
  \parbox{120pt}{\begin{fmfgraph*}(120,60)
    \fmfkeep{planar-3}
    \fmfleft{w}
    \fmfright{e}
    \fmf{dashes_arrow,label=$p$}{w,cw}
    \fmf{dashes_arrow,right,tension=0.5,label=$k$}{ce,cw}
    \fmf{plain_arrow,right,tension=0.5,label=$k+p$}{cw,ce}
    \fmf{dashes_arrow}{ce,e}
    \fmfdot{cw,ce}
    \fmfv{label=$i$,label.angle=60}{w}
    \fmfv{label=$j$,label.angle=120}{e}
  \end{fmfgraph*}} + \parbox{90pt}{\fmfreuse{pion-counterterm}}
  \label{comm-cancel} 
 \eea 
\end{fmffile}
%
Because the full noncommutative calculation we are about to explore 
subsumes this as the special case where $\theta\rightarrow 0$, we 
will not do this calculation explicitly: for a demonstration using 
dimensional regularization, the reader is referred to \cite{ps}.  We 
do point out that the counterterm displayed is not arbitrary, but 
completely fixed (at zero external momentum) by the renormalization 
condition on the sigma tadpole, as inspection of (\ref{counterterms}) 
reveals.  At one loop the sigma tadpole counterterm is fixed by the 
requirement that
\begin{fmffile}{tadpoles}
 \bea 
  \parbox{90pt}{\begin{fmfgraph*}(90,50)
      \fmfkeep{tadpole-1}
      \fmfleft{w}
      \fmfright{e}
      \fmf{plain,tension=2}{w,c}
      \fmf{plain_arrow,left,label=$k$}{c,e}
      \fmf{plain,right}{c,e}
      \fmfdot{c}
  \end{fmfgraph*}}
  +
  \parbox{90pt}{\begin{fmfgraph*}(90,50)
      \fmfkeep{tadpole-2}
      \fmfleft{w}
      \fmfright{e}
      \fmf{plain,tension=2}{w,c}
      \fmf{dashes_arrow,left,label=$k$}{c,e}
      \fmf{dashes,right}{c,e}
      \fmfdot{c}
  \end{fmfgraph*}}
  +
  \parbox{50pt}{\fmfreuse{sigma-tadpole-counterterm}}
  = 0 
 \label{tadpole-renorm}
 \eea 
\end{fmffile}
%
Now consider the noncommutative case.  The first, perhaps surprising, fact is 
that noncommutativity does {\it not} alter the tadpole graphs at 
one-loop. This follows from the elementary observation that while the one-loop 
tadpole graphs above contain cubic vertices (and so yield nontrivial 
noncommutative phase terms), momentum conservation dictates that no 
external momentum flows into the cubic vertices, so that phase factor 
(\ref{gen-phase}) always degenerates to $1$, or equivalently the cosine terms 
in the symmetrized vertices always vanish.

Thus the sigma tadpole counterterm will not be modified in the 
noncommutative case.  Nonetheless we need to explicitly calculate 
it from (\ref{tadpole-renorm}).  Since loop integrals containing 
phases like (\ref{gen-phase}) are most easily evaluated using 
Schwinger parameters (see for example \cite{iz},\cite{z}), we will 
henceforth use them, coupled with a $C^{\infty}$ smooth UV momentum 
cutoff regularization throughout.  The basic Schwinger parameter 
representation is:
\be
{1\over k^{2} + m^{2}} = \int_{0}^{\infty} d\alpha 
e^{-\alpha(k^{2}+m^{2})}      
\ee
%
whence
\be
\int {d^{4}k\over k^{2}+m^{2}} = \int d\Omega_{4} \int_{0}^{\infty} 
d\alpha \int dk k^{3} e^{-\alpha (k^{2}+m^{2}) }
= \pi^{2} \int_{0}^{\infty} {d\alpha\over \alpha^{2}} e^{-\alpha^{2} m^{2}}  
\ee
%
Thus high momentum divergences are replaced with small $\alpha$ 
divergences which are regulated by inserting a factor of $e^{-1/(\alpha 
\Lambda^{2})}$ under the $\alpha$ integral, where $\Lambda$ is the 
(fundamental) UV momentum cutoff.  The basic integral we will 
encounter repeatedly is then of the form:
\be
\int^{\infty}_{0} {d\alpha\over\alpha^{2}} e^{-\alpha m^{2} - 
{x^{2}\over \alpha m^{2}} }  =  {m^{2}\over x^{2}} + m^{2} 
\log(x^{2}) + m^{2} (2 \gamma -1) + O(x)     
\ee
%
in the limit of small $x(= {m\over\Lambda})$. This is discussed and 
derived in the appendix.  Thus:
\bea
\parbox{90pt}{\fmfreuse{tadpole-1}} & = & {1\over 2} (-6 v\lambda) 
\int {d^{4}k\over (2\pi^{4})} {1\over k^{2} + m_{\sigma}^{2}} 
\nonumber \\
& = & - {3 \lambda v\over 16 \pi^{2}} \left[ \Lambda^{2} - 
m_{\sigma}^{2} \log({\Lambda^{2}\over m_{\sigma}^{2}}) \right. \nonumber \\
& & \qquad \qquad \left. + m_{\sigma}^{2}(2\gamma -1) + O({m_{\sigma}\over 
\Lambda}) \right]           
\eea
%
where $m_{\sigma}^{2} \equiv 2\mu^{2}$ is the (tree-level) mass-squared 
for the sigma.  Next,
\bea
\parbox{90pt}{\fmfreuse{tadpole-2}} & = & {1\over 2} (-2 v\lambda 
\delta^{ij}) \int {d^{4}k\over (2\pi^{4})} {\delta^{ij} \over k^{2} + \xi^{2}} 
\nonumber \\
& = & - {(N-1) \lambda v\over 16 \pi^{2}} \left[ \Lambda^{2} - 
\xi^{2} \log({\Lambda^{2}\over \xi^{2}}) \right. \nonumber \\
& & \qquad \qquad \left. + \xi^{2}(2\gamma -1) + O({\xi\over 
\Lambda}) \right]           
\eea
%
where $\xi$ is a small infrared mass for the pion, to be taken to zero 
later.  Thus (\ref{tadpole-renorm}) fixes the tadpole counterterm to 
be:
\bea
\parbox{50pt}{\fmfreuse{sigma-tadpole-counterterm}} & = & -(\delta_{\mu}v + 
\delta_{\lambda} v^{3}) \nonumber \\
&=& {\lambda v \over 16 \pi^{2}} \left\{ \Lambda^{2} (N+2) - 3 
m_{\sigma}^{2} \log({\Lambda^{2}\over m_{\sigma}^{2}}) - \xi^{2}(N-1) 
\log({\Lambda^{2}\over\xi^{2}}) \right. \nonumber \\
& & \left. + (2\gamma -1)\left[ 3m_{\sigma}^{2}+ \xi^{2}(N-1)\right] 
+ O({1\over \Lambda}) \right\}  
\eea
%
As discussed this fixes the pion propagator counterterm at zero 
momentum (and at one loop).  Thus, including the momentum-dependent wave 
function renormalization term we have:
\bea
\parbox{90pt}{\fmfreuse{pion-counterterm}} = 
&=& {\lambda \delta^{ij} \over 16 \pi^{2}} \left\{ \Lambda^{2} (N+2) - 3 
m_{\sigma}^{2} \log({\Lambda^{2}\over m_{\sigma}^{2}}) - \xi^{2}(N-1) 
\log({\Lambda^{2}\over\xi^{2}}) \right. \nonumber \\
& & \left. + (2\gamma -1)\left[ 3m_{\sigma}^{2}+ \xi^{2}(N-1)\right] 
+ O({1\over \Lambda}) \right\} + \delta^{ij} \delta_{z} p^{2}
\label{pion-counterterm}
\eea
%
Now consider the one-loop contributions to the pion propagator.  In order to
consider separately the (noncommuting) \cite{mrs} UV ($\Lambda \rightarrow 
\infty$), and IR ($p \rightarrow 0$) limits, we will not set $p=0$ 
{\it a priori}, but will compute the $\pi\pi$ amplitude $\Gamma^{(2)}_{1PI}$ 
for an arbitary external momentum $p$.  The first graph 
in (\ref{comm-cancel}) has a single quartic vertex (\ref{four-pi}).  Denoting
the loop momentum by $k$, (\ref{four-pi}) yields a phase factor of:
\bea
& &f \left(2 \cos^{2}({p\times k\over 2}) + N-1 \right) + (1-f)\Bigg( 2 + 
(N-1) \cos(p\times k)\Bigg) \nonumber \\ 
& & = fN + 2(1-f) + \bigg( f + (1-f)(N-1) \bigg) \cos(p\times k)
\eea
The constant term is the planar contribution to the noncommutative analogue
of the first graph in (\ref{comm-cancel}), and is explicitly:
\bea
\parbox{120pt}{\fmfreuse{planar-1}} &=& {1\over 2} \left[-2 \delta^{ij} 
\lambda (f N + 2 (1-f) ) \right] \int {d^{4}k\over (2\pi)^{4}} 
{1\over k^{2}+\xi^{2}} \nonumber \\
&=& - {\lambda \delta^{ij} \left[ fN + 2(1-f) \right] \over 16 \pi^{2}} 
\left[ \Lambda^{2} - \xi^{2} \log(\Lambda^{2}/\xi^{2}) + \right. \nonumber 
\\ & & \qquad \left. \xi^{2} (2\gamma-1) + O({\xi\over \Lambda}) \right]
\label{planar-1}
\eea
%
whereas writing $\cos(k\times p)= {1\over 2} (e^{i k\times p} + 
e^{i p\times k})$, and using the fact that the exponentials are the same
under the integral over all momenta, the nonplanar contribution is:
\begin{fmffile}{nonplanar}
\bea
  \parbox{120pt}{\begin{fmfgraph*}(120,60)
     \fmfleft{w}
     \fmfright{e}
     \fmfv{label=$i$,label.angle=60}{w}
     \fmfv{label=$j$,label.angle=120}{e}
     \fmf{dashes_arrow,label=$p$}{w,cw}
     \fmf{phantom,tension=0.2,tag=1}{cw,ce}
     \fmf{phantom,tag=2}{ce,e}
     \fmf{dashes_arrow,left,tension=0.2,label=$k$}{cw,ce}
     \fmf{dashes,right,tension=0.2}{cw,ce}
     \fmfdot{cw}
     \fmfposition
     \fmfipath{p[]}
     \fmfiset{p1}{vpath1(__cw,__ce)}
     \fmfiset{p2}{vpath2(__ce,__e)}
     \fmfi{dashes}{subpath (0,14/15*length(p1)) of p1}
     \fmfi{dashes}{subpath (1/10*length(p2),length(p2)) of p2}
  \end{fmfgraph*}}
&=& {1\over 2} \left[- 2 \lambda \delta^{ij} (f+(1-f)(N-1)) \right] \int {d^{4}k\over 
(2\pi)^{4}} {e^{i k\times p} \over k^{2}+\xi^{2}} \nonumber \\
&=& - {\lambda \delta^{ij} \left[f+(1-f)(N-1)\right] \over 16 \pi^{2}} \left[ \Lambda_{eff}^{2} - 
\xi^{2} \log(\Lambda_{eff}^{2}/\xi^{2}) + \right. \nonumber \\ & & 
\qquad \left. \xi^{2} (2\gamma-1) + O({\xi\over \Lambda_{eff}}) \right] 
\label{nonplanar-1}
\eea  
%
where
\be
\Lambda_{eff}^{2} = {1\over {1\over \Lambda^{2}} + p\circ p}
\ee
%
and 
\be
p \circ q \equiv - {p_{\mu} \theta^{2}_{\mu\nu} q_{\nu} \over 4}   
\ee
%
in a basis where $\theta^{\mu\nu}$ is skew-symmetric (the coordinates 
form pairs of noncommuting coordinates), so that its square is 
diagonal.  This effective cutoff arises because completing the 
square in the Schwinger integral now yields an inhomogeneous term 
that is the same form as the fundamental UV cutoff regulator:
\bea
{e^{i k\times p} \over k^{2}+ m^{2}} &=&  \int_{0}^{\infty} d\alpha 
e^{-\alpha (k^{2} + m^{2}) + i k\times p} \nonumber \\
&=&  \int_{0}^{\infty} d\alpha e^{-\alpha (l^{2}+ m^{2})} 
e^{-{1\over 4 \alpha} p_{\rho} p^{\nu} \theta_{\mu\nu} 
\theta^{\mu\rho}} \nonumber \\
&\rightarrow_{reg}& \int_{0}^{\infty} d\alpha e^{-\alpha (l^{2} + 
m^{2}) - {1\over \alpha \Lambda_{eff}^{2}} }      
\eea
%
where $l_{\mu} = k_{\mu} - i/(2\alpha) \theta_{\mu\nu} p^{\nu}$ is a 
linear change of variables with respect to the $k$ integral.  Note 
the (irrelevant) $1/4$ factor in the definition of $p\circ q$ which 
does not appear in \cite{mrs}.

The second diagram in (\ref{comm-cancel}) is handled similarly.  The phase
factor from (\ref{pi-pi-sigma-sigma}) is now:
\be
f + (1-f) \cos(k\times p)
\ee
Thus the planar contribution to the noncommutative analogue of the second 
graph in (\ref{comm-cancel}) is:
\bea
\parbox{120pt}{\fmfreuse{planar-2}} &=& {1\over 2} (-2 \lambda \delta^{ij} f) 
\int {d^{4}k\over (2\pi)^{4}} {1\over k^{2} + m_{\sigma}^{2}} \nonumber \\
&=& - {\lambda \delta^{ij} f\over 16 \pi^{2}} \left[ \Lambda^{2} - 
m_{\sigma}^{2} \log(\Lambda^{2}/m_{\sigma}^{2}) + \right. \nonumber \\ & & 
\qquad \left. m_{\sigma}^{2} (2\gamma-1) + O({m_{\sigma} \over \Lambda}) \right]
\label{planar-2}
\eea
and the nonplanar contribution is:
\bea
  \parbox{120pt}{\begin{fmfgraph*}(120,60)
     \fmfleft{w}
     \fmfright{e}
     \fmfv{label=$i$,label.angle=60}{w}
     \fmfv{label=$j$,label.angle=120}{e}
     \fmf{dashes_arrow,label=$p$}{w,cw}
     \fmf{phantom,tension=0.2,tag=1}{cw,ce}
     \fmf{phantom,tag=2}{ce,e}
     \fmf{plain_arrow,left,tension=0.2,label=$k$}{cw,ce}
     \fmf{plain,right,tension=0.2}{cw,ce}
     \fmfdot{cw}
     \fmfposition
     \fmfipath{p[]}
     \fmfiset{p1}{vpath1(__cw,__ce)}
     \fmfiset{p2}{vpath2(__ce,__e)}
     \fmfi{dashes}{subpath (0,14/15*length(p1)) of p1}
     \fmfi{dashes}{subpath (1/10*length(p2),length(p2)) of p2}
   \end{fmfgraph*}}
&=& -\lambda \delta^{ij} (1-f) \int {d^{4}k\over (2\pi)^{4}}
{e^{i k\times p} \over k^{2}+m_{\sigma}^{2}} \nonumber \\
&=& - {\lambda \delta^{ij} (1-f) \over 16 \pi^{2}} \left[ \Lambda_{eff}^{2} - 
m_{\sigma}^{2} \log(\Lambda_{eff}^{2}/m_{\sigma}^{2}) + \right. 
 \nonumber \\ & & 
\qquad \left. m_{\sigma}^{2} (2\gamma-1) + O({m_{\sigma} \over 
\Lambda_{eff}}) \right] 
\label{nonplanar-2}
\eea
%
The third diagram of (\ref{comm-cancel}) will yield momentum dependent 
corrections to the propagator, as the external momentum circulates in the 
loop.  Note also that it does not depend on $f$ since there are no quartic 
vertices.  Instead of using two Schwinger parameters for the two internal 
propagators, we will combine the propagators using a Feynman parameter, and 
then use a Schwinger parameter via the identity:
\be
{1\over \left( l^{2}+ \Delta^{2}\right)^{2}} = -{d\over d (l^{2})} 
\left[ \int_{0}^{\infty} d\alpha e^{-\alpha (l^{2} + \Delta^{2})} 
\right] = \int_{0}^{\infty} d\alpha \alpha e^{-\alpha (l^{2} + \Delta^{2}) }
\ee
%
Then we will need (see appendix) the small x expansion:
\be
\int {d\alpha\over \alpha} e^{-\alpha \Delta^{2} - {x^{2}\over 
\alpha \Delta^{2}} } = - \log(x^{2}) - 2 \gamma + O(x)    
\ee
%
which makes explicit the logarithmic (as opposed to quadratic) divergence
of this type of diagram.  Now, since we have two cubic vertices, we pick up 
two phase factors due to noncommutativity.  Denoting by $k$ the pion momentum 
in the loop, the vertex (\ref{pi-pi-sigma}) yields for the third
diagram of (\ref{comm-cancel}) the phase:
\be
\cos({k\times p\over 2})\cos({-k\times -p\over 2}) = {1\over 2} \left( 1 + 
\cos({k\times p}) \right) 
\ee
%
Thus planar and nonplanar contributions are weighted equally.  The former
yields
\bea
\parbox{120pt}{\fmfreuse{planar-3}} &=& {1\over 2} 4 v^{2} 
\lambda^{2} \delta^{ij} \int {d^{4}k\over (2\pi)^{4}} {1\over 
k^{2}+ \xi^{2}} {1\over (k+p)^{2} + m_{\sigma}^{2}} \nonumber        
\eea
\bea
&=& 2v^{2}\lambda^{2}\delta^{ij} \int_{0}^{1} dx \int {d^{4}l\over 
(2\pi)^{4}} {1\over (l^{2}+\Delta^{2})^{2}} \nonumber \\
&\rightarrow_{reg}& {v^{2}\lambda^{2}\delta^{ij}\over 8\pi^{2}} 
\int_{0}^{1} dx \int_{0}^{\infty} {d\alpha\over\alpha} e^{-\alpha \Delta^{2} - 
{1\over\alpha \Lambda^{2}} } \nonumber \\
&=& {-v^{2}\lambda^{2}\delta^{ij} \over 8\pi^{2}}  \left[ 
\int_{0}^{1} \log ({\Delta^{2} \over \Lambda^{2}}) dx + 2\gamma + 
O({\Delta\over\Lambda}) \right]   
\label{planar-3}
\eea
%
where $\Delta^{2} = x(1-x)p^{2} + x \xi^{2} + (1-x) m_{\sigma}^{2}$, 
$l= k + (1-x)p$, and where in the second line we introduced the 
Feynman parametrization:
\be
{1\over ( k^{2} + \xi^{2}) [(k+p)^{2}+m_{\sigma}^{2}])} = \int_{0}^{1} 
dx {1\over ( l^{2} + \Delta^{2} )^{2}  }    
\ee
%
To proceed, we can take the IR regulator $\xi$ to zero without any 
difficulty, to simplify the Feynman parameter integral, so that 
(recalling $v^{2} = \mu^{2}/\lambda$) 
\be
(\ref{planar-3}) = {-\lambda m_{\sigma}^{2}\delta^{ij} \over 
16\pi^{2}} \left[ \log \left({p^{2}+m_{\sigma}^{2}\over 
\Lambda^{2}} \right) + 
{m_{\sigma}^{2}\over p^{2}} \log\left({p^{2}+m_{\sigma}^{2}\over 
m_{\sigma}^{2}}\right) + 2 (\gamma-1) + O({\Delta\over \Lambda}) 
\right]        
\ee
%
Noting that $l=k+(1-x)p$ implies that $k\times p = l\times p$, and writing
as usual $\cos(k\times p) = \cos(l\times p)$ in exponential form, and noting 
the symmetry of the terms under integration over all $l$, the 
nonplanar contribution to the third graph in (\ref{comm-cancel}) is:
\bea
  \parbox{120pt}{\begin{fmfgraph*}(120,60)
     \fmfleft{w}
     \fmfright{e}
     \fmfv{label=$i$,label.angle=60}{w}
     \fmfv{label=$j$,label.angle=120}{e}
     \fmf{dashes_arrow,label=$p$}{w,cw}
     \fmf{dashes}{ce,e}
     \fmf{plain_arrow,left=0.7,tension=0.3,label=$k+p$}{cw,c}
     \fmf{plain,right=0.7,tension=0.3}{c,ce} 
     \fmf{dashes_arrow,left=0.7,tension=0.3,rubout,label=$k$}{c,cw}
     \fmf{dashes,left=0.7,tension=0.3,rubout}{c,ce}
     \fmfdot{cw,ce}
  \end{fmfgraph*}}
&=& 2 v^{2}\lambda^{2}\delta^{ij} \int {d^{4}k\over (2\pi)^{4}} {1\over 
k^{2}+ \xi^{2}} {e^{i k\times p} \over (k+p)^{2} + m_{\sigma}^{2}} 
\nonumber \\      
&=& 2 v^{2}\lambda^{2}\delta^{ij} \int_{0}^{1} dx \int {d^{4}l\over 
(2\pi)^{4}} {e^{i l\times p}\over (l^{2}+\Delta^{2})^{2}} \nonumber    
\eea
\be
= {-\lambda m_{\sigma}^{2}\delta^{ij} \over 
16\pi^{2}} \left[ \log\left({p^{2}+m_{\sigma}^{2}\over 
\Lambda_{eff}^{2}}\right) + 
{m_{\sigma}^{2}\over p^{2}} \log\left({p^{2}+m_{\sigma}^{2}\over 
m_{\sigma}^{2}}\right) + 2 (\gamma-1) + O({\Delta\over \Lambda_{eff} }) 
\right]        
\label{nonplanar-3}
\ee
\end{fmffile}
%
Thus the effect of noncommutativity is to re-weight the planar and 
nonplanar graphs with respect to the commutative graphs (where there is no 
distinction between planarity and nonplanarity) in the cases where nonplanar
graphs are generated, and replace the $\Lambda$ in the 
planar graphs with $\Lambda_{eff}$ in the nonplanar graphs.  The one-loop 
correction to the (inverse) propagator is the sum 
of the six graphs (\ref{planar-1}), (\ref{nonplanar-1}), (\ref{planar-2}), 
(\ref{nonplanar-2}), (\ref{planar-3}), (\ref{nonplanar-3}), and the 
counterterm (\ref{pion-counterterm}).  This sum is the noncommutative 
equivalent of (\ref{comm-cancel}).  Explicitly it is equal to:
\bea
\sum_{1-loop} &=& {\lambda \delta^{ij}\over 16\pi^2} \left\{ \bigg[ N (1-f)
+ f \bigg] \bigg(\Lambda^{2}- \Lambda_{eff}^2 \bigg) + \bigg(2-f \bigg) 
m_{\sigma}^{2} \log({\Lambda_{eff}^{2}\over \Lambda^{2}}) \right. \nonumber \\ 
& & \qquad \qquad \left. + 2 m_{\sigma}^{2} \left( 1- {p^{2} + m_{\sigma}^{2}
\over p^{2}} \log({p^{2}+m_{\sigma}^{2}\over m_{\sigma}^2}) 
\right) \right\}] + \delta^{ij} \delta_{Z} p^{2}
\eea
%
or eliminating $\Lambda_{eff}$ in favour of $\Lambda$, $\theta$, and $p$:
\bea
\sum_{1-loop} &=& {\lambda \delta^{ij}\over 16\pi^2} \left\{ \bigg[ N (1-f) 
+ f \bigg] \Lambda^{2} \left( 1- {1\over 1 + \Lambda^{2} (p \circ p)} 
\right) - \bigg(2-f \bigg) m_{\sigma}^{2} \log(1+\Lambda^{2} (p\circ p)) 
\right. \nonumber \\ 
& & \left. \qquad \qquad + 2 m_{\sigma}^{2} \left( 1- {p^{2} + m_{\sigma}^{2}
\over p^{2}} \log({p^{2}+m_{\sigma}^{2}\over m_{\sigma}^2}) \right) 
\right\} + \delta^{ij} \delta_{Z} p^{2}
\label{big-result}
\eea
\section{Discussion}

This result is rather remarkable, and some comments are in order.  First note 
that if we take the noncommutativity $\theta$ to zero, the explicit 
dependence on $\Lambda$ cancels, and we are left with the usual finite 
commutative result:
\be
\sum_{1-loop,comm} = {\lambda \delta^{ij}\over 8 \pi^2} m_{\sigma}^{2} 
\left( 1- {p^{2} + m_{\sigma}^{2} \over p^{2}} \log({p^{2}+m_{\sigma}^{2}
\over m_{\sigma}^2}) \right) + \delta^{ij} \delta_{Z} p^{2}
\ee
%
which furthermore vanishes in the $p\rightarrow 0$ limit. Thus the mass shift
of the pions due to one-loop quantum corrections vanishes; this is just 
Goldstone's theorem.  However, for a strictly nonzero $\theta$, the result
is unavoidably UV cutoff dependent (the freedom to choose $\delta_{Z}$ clearly
does not help us because the divergences are not proportional to $p^2$), and 
therefore naively divergent (both quadratically and logarithmically) in the 
continuum ($\Lambda\rightarrow \infty$) limit.  This cannot be evaded by any
choice of $f \in [0,1]$, or N.  However, viewed as a Wilsonian action, for 
which $\Lambda$ is a fixed, physical parameter, the limit $p\rightarrow 0$ 
exists, and (\ref{big-result}) becomes zero: again yielding Goldstone's 
theorem in this ordering of the limits. The conclusion is that not only do UV 
and IR limits not commute, but the continuum limit of the renormalized theory 
is inconsistent with Nambu-Goldstone realization of the global symmetry.  We 
emphasize, that this occurs after the renormalization programme has been 
carried out; we have used up the counterterm freedom in fixing the sigma 
vev.

This is not an artifact of the renormalization prescription (of which
we have only used the condition imposing nonrenormalization of the sigma 
vev), since clearly if we somehow arrange the $\pi\pi$ amplitude to contain 
no net fundamental $\Lambda$ dependence, post-renormalization $\Lambda$ 
dependence will be shifted into the sigma tadpole.  This is a direct 
consequence of the restrictive counterterm structure present in a 
spontaneously broken theory.

To summarize, because no external momentum flows into a tadpole vertex, 
noncommutativity does not affect tadpole graphs, and hence the sigma tadpole
renormalization counterterm is not modified.  Since tadpole terms reflect 
shifts in vacuum expectation values of fields, and since the sigma vev 
measures the spontaneous symmetry breaking of the theory, colloquially, 
spontaneous symmetry breaking is blind to the noncommutativity at this level. 
On the other hand, since the counterterm structure is fixed, this means that 
the pion propagator counterterm is not modified (modulo wave function 
renormalization).  However, the graphs that contribute to the one-loop 
quantum corrections of the pion propagator are modified, and the usual 
cancellation that ensures the pion masses are finite (and zero) after quantum 
corrections is violated in such a way, that cutoff dependence is restored; 
so UV and IR limits do not commute.  In sum, we conclude that the 
Nambu-Goldstone realization of the $O(N)$ symmetry of this model is not 
perturbatively compatible with the continuum renormalization of the 
noncommutative theory.  The interpretation and implications of this result 
will be explored elsewhere.  

\section*{Acknowledgements}

This work was supported in part by the Natural Sciences and 
Engineering Research Council of Canada.  KK would like to thank R.
Allahverdi for useful discussions, and D. Shaw for help with the Feynmf
package.


\appendix

\section{Appendix}

\subsection{The Feynman Rule for Noncommutative Phases}

Proceeding inductively, we have:
\bea
& &((\phi_{1}\ast ...\phi_{n})\ast \phi_{n+1}) (x) \nonumber \\ 
&=&\int \prod_i {d^{4}p_{i}\over (2\pi)^2} \lim_{y,z\rightarrow x} e^{{i\over 2} \theta^{\mu\nu} \partial_{\mu}^{y} \partial_{\nu}^{z}} \left( 
e^{i (p_{1}+\ldots+ p_{n})\cdot y} e^{i p_{n+1}\cdot z} \right) 
e^{-{i\over 2} \sum_{i<j}^{n} p_{i} \times p_{j}} \tilde{\phi}_{1}(p_{1}) 
\ldots \tilde{\phi}_{n+1} (p_{n+1}) \nonumber \\ 
&=& \int \prod_i {d^{4}p_{i}\over (2\pi)^2} e^{-{i\over 2} \theta^{\mu\nu} (p_{1}+\ldots
p_{n})_{\mu} (p_{n+1})_{\nu}} e^{i (p_{1}+\ldots+ p_{n+1})
\cdot x}  e^{-{i\over 2} \sum_{i<j}^{n} 
p_{i} \times p_{j}} \tilde{\phi}_{1}(p_{1}) \ldots \tilde{\phi}_{n+1}
(p_{n+1}) \nonumber \\
&=& \int \prod_i {d^{4}p_{i}\over (2\pi)^2} 
e^{i (p_{1}+\ldots+ p_{n+1}) \cdot x} 
e^{-{i\over 2} \sum_{i<j}^{n+1} p_{i} \times p_{j}} 
\tilde{\phi}_{1}(p_{1}) \ldots \tilde{\phi}_{n+1}(p_{n+1})
\eea
%
where the induction hypothesis is used in the second line.  Thus,
\be
V(p_{1},...,p_{n})= e^{-{i\over 2} \sum_{i<j} p_{i}\times p_{j}}.
\ee

\subsection{A Useful Integral}

In evaluating the loop integrals in this paper, we encountered integrals of 
the form 
\be
{\cal I}_{n}(x) \equiv \int_{0}^{\infty} {d\alpha\over\alpha^{n}} e^{-\alpha 
m^{2}- {x^{2}\over \alpha m^{2}}}
\ee
%
which can be taken as definitions (modulo constants) of the $K$ type modified 
Bessel functions.  Since we are really only interested in the small $x$ 
(large $\Lambda$) behaviour, and in the cases $n=1,2$, an elementary treatment 
suffices to determine dominant behaviour.  We split the integration region 
into $[0,x/m^{2}]$, and $[x/m^{2},\infty)$, and then expand the appropriate
exponential on each region.  For $n=2$, after changing variables $\alpha 
\rightarrow 1/\alpha \cdot x^{2}/m^{4}$
\bea
{\cal I}_{2}(x) &=& {m^{4}\over x^{2}} \int_{0}^{\infty} e^{-\alpha m^{2} - 
{x^{2}\over \alpha m^{2}}} d\alpha \nonumber \\
&=& {m^{4}\over x^{2}} \left[ \int_{0}^{x\over m^{2}} \left( 1 - \alpha m^{2}
+ {\alpha^{2} m^{4}\over 2} + O(\alpha^{3}) \right) \cdot e^{-{x^{2}\over m^{2}
\alpha}} \quad d\alpha  \right. \nonumber \\
& & \qquad + \left. \int_{x\over m^{2}}^{\infty} \left( 1- {x^{2}\over m^{2}
\alpha} + O(x^{4}) \right) e^{-\alpha m^2} d\alpha \right] \nonumber \\
&=& {m^{2}\over x^{2}} \left[ (x-{x^{2}\over 2}) e^{-x} - x^{2} 
\textnormal{Ei}_{1}(x) + O(x^{3}) + e^{-x} - x^{2} \textnormal{Ei}_{1}(x) 
+ O(x^{4}) \right] \nonumber \\
&=& {m^{2}\over x^{2}} \left[ 1-x^{2} + 2 x^{2} \log(x) + 2 x^{2} \gamma + 
O(x^{3}) \right] \nonumber \\
&=& {m^{2}\over x^{2}} + m^{2} \log(x^{2}) + m^{2} (2\gamma -1) + O(x)
\eea
%
where we have used the exponential integral
\be
\textnormal{Ei}_{1}(x) = \int_{1}^{\infty} {e^{-xt}\over t} dt
\ee
%
and its asymptotic expansion near $x=0$, discovered through:
\bea
{d\over dx}\left[ \textnormal{Ei}_{1}(x) \right] &=& -\int_{1}^{\infty}
e^{-xt} dt = - {e^{-x}\over x} \nonumber \\
&=& - {1\over x} + \sum_{n=1}^{\infty} {(-x)^{n-1}\over n!} \nonumber \\
\eea
%
whence
\be
\textnormal{Ei}_{1}(x) = -\log(x) + C + O(x)
\ee
%
It can be shown that $C=-\gamma$, but since it must cancel out of all
physical amplitudes, we heed it no further. The $n=1$ case is handled
identically, and we omit it here. 


\end{document}